\documentclass[twocolumn]{aastex631}
\usepackage{amsmath}
\usepackage{comment}
\usepackage{booktabs}
\usepackage{breqn}

\newcommand{\deldel}[2]{\frac{\partial #1}{\partial #2}}
\newcommand{\sign}[1]{{\rm{}sign}(#1)}
\newcommand{\bex}{\hat{\mathbf{e}}_x}
\newcommand{\bey}{\hat{\mathbf{e}}_y}
\newcommand{\bez}{\hat{\mathbf{e}}_z}
\newcommand{\bno}{\hat{\mathbf{n}}_{\rm{}o}}
\newcommand{\bni}{\hat{\mathbf{n}}_{\rm{}i}}
\newcommand{\blo}{\hat{\mathbf{l}}_{\rm{}o}}
\newcommand{\bli}{\hat{\mathbf{l}}_{\rm{}i}}
\newcommand{\bmo}{\hat{\mathbf{m}}_{\rm{}o}}
\newcommand{\bmi}{\hat{\mathbf{m}}_{\rm{}i}}
\newcommand{\Co}{\mathbf{c}_{\rm{o}}}
\newcommand{\Cox}{c_{{\rm o}x}}
\newcommand{\Coy}{c_{{\rm o}y}}
\newcommand{\Coz}{c_{{\rm o}z}}
\newcommand{\Ci}{\mathbf{c}_{\rm{i}}}
\newcommand{\Cix}{c_{{\rm i}x}}
\newcommand{\Ciy}{c_{{\rm i}y}}
\newcommand{\Ciz}{c_{{\rm i}z}}
\newcommand{\So}{\mathbf{S}_{\rm{}o}}
\newcommand{\Si}{\mathbf{S}_{\rm{}i}}

\newcommand{\phii}{\phi_{\rm{i}}}
\newcommand{\phio}{\phi_{\rm{o}}}
\newcommand{\ii}{i_{\rm{i}}}
\newcommand{\io}{i_{\rm{o}}}
\newcommand{\pai}{{\rm{PA_i}}}
\newcommand{\pao}{{\rm{PA_o}}}

\newcommand{\Ns}{\mathbf{N}_{\rm{S}}}
\newcommand{\dc}{\delta\mathbf{c}}
\newcommand{\lr}[1]{\left( #1 \right)}

\newcommand{\chtext}[1]{{\color{black}{#1}}}

\begin{document}

\title{Shadow-Based Framework for Estimating Transition Disk Geometries}

\author[0000-0003-4039-8933]{Ryuta Orihara}
\affiliation{College of Science, Ibaraki University, 2-1-1 Bunkyo, Mito, Ibaraki 310-8512, Japan}
\affiliation{Department of Astronomy, Graduate School of Science, The University of Tokyo, 7-3-1 Hongo, Bunkyo-ku, Tokyo 113-0033, Japan}
\author[0000-0002-3001-0897]{Munetake Momose}
\affiliation{College of Science, Ibaraki University, 2-1-1 Bunkyo, Mito, Ibaraki 310-8512, Japan}



\begin{abstract}
Some transition disks host misaligned inner disks with radii of several au. Understanding the geometric and physical properties of these misaligned disks is essential for advancing terrestrial planet formation models. This study introduces a novel method to infer the three-dimensional structures of both inner and outer disks by analyzing non-axisymmetric shadows and the horizon in optical and infrared scattered light images of the outer disk. This method was applied to the HD 100453 system, in which infrared scattered light images from the Very Large Telescope revealed disk shadows. These results indicate that the inner disk is misaligned by $\sim70^\circ$ relative to the outer disk, which is consistent with the results of previous studies. \chtext{The aspect ratio of the inner disk surface was estimated to be 0.17, which may reflect the surface height of the optically thick dusty component due to vertical lofting by MHD winds or turbulence.} In addition, the surface height distribution of the outer disk was characterized, providing novel insights into its vertical structure. 
\end{abstract}

\keywords{Protoplanetary disks (1300), Infrared astronomy (786), Optical astronomy (1776)}


\section{Introduction} \label{sec:introduction}

    Recent observations at radio, optical, and infrared (OIR) wavelengths have revealed a wide variety of structures within protoplanetary disks that surround young stars \citep{Benisty_2023, Pinte_2023}. Among these protoplanetary disks, those with an inner cavity resulting from dust depletion in the inner region are called transition disks \citep{Espaillat_2014}.
    Observations of dust continuum emissions using the Atacama Large Millimeter/submillimeter Array (ALMA) have provided insight into the density and temperature distributions within these disks, and the presence of compact dust disks within the cavities of several transition disks has been confirmed. Research by \cite{Francis_2020} indicated that approximately half of the observed sources have an inner disk, suggesting that dust in the innermost regions of disks is either persistent or sustained by a dust supply rate comparable to the stellar accretion rate. 
    
    Although compact inner disks have been detected, only a limited number of targets were successfully spatially resolved at a resolution sufficient to reveal their geometries. Attempts have been made to observe the shape of the inner disk directly with the Very Large Telescope Interferometer (VLTI) \citep{Bohn_2022}, but these are based on parametric models of visibilities that are required for additional components other than the inner disk, making it difficult to discuss uncertainties in the derived disk parameters. In addition, the use of an interferometer combining four telescopes resulted in a smaller data sample in the visibility plane compared to that from ALMA. Therefore, analysis of the inner disk structure based on imaging remains challenging with current instrumentation. 
    
    The geometric structure and physical properties of inner disks can be estimated from existing data by exploiting the effect of the inner disk on the outer disk. A notable observational feature in this context is the shadow cast within the outer disks by a misaligned inner disk. 
    \chtext{Examples of such systems include HD100453 \citep{Benisty_2017}, HD142527 \citep{Marino_2015}, CQ Tau \citep{Uyama_2020}, DoAr44 \citep{Casassus_2018}, and J1604 \citep{Mayama_2012}. 
    The misalignment of the inner disk relative to the outer disk can be caused by several mechanisms, including gravitational perturbations from a planet or a stellar companion \citep{Nealon_2018, Zhu_2019}, warping induced by a misaligned stellar magnetic field, \citep{Bouvier_1999}, or anisotropic accretion of material onto the disk \citep{Bate_2018}. 
    }

    \chtext{
    A shadow cast on the outer disk can also have a significant effect on its evolution. According to \cite{Montesinos_2016} and \cite{Cuello_2019}, the local cooling caused by the shadow of the inner disk can potentially trigger the formation of spiral arms in the outer disk. A localized temperature drop induces an azimuthal pressure gradient. As the gas rotates and moves through the shadowed region, it undergoes azimuthal acceleration, leading to the development of spiral arms. Over time, these spirals can evolve into ring-like features, as demonstrated by numerical simulations \citep{Ziampras_2024, Su_2024, Zhang_2024}. This process can occur even in gravitationally stable disks without planets, indicating the potential for planetesimal formation over a wide range of evolutionary stages.
    }
    
    A framework that links the shadow to the inner disk orientation was proposed by \cite{Min_2017}. In this analytical model, the two shadows on the outer disk are calculated under the given position angle (PA) and inclination ($i$) of both the inner and outer disks, and the height of the scattering surface ($h$) of the outer disk. However, this method has several limitations in estimating the shape of the inner disk from its shadow. One such limitation is the neglect of the thickness of the inner disk, which results in infinitesimally narrow shadows on the disk. Because an actual shadow has a finite width, the model accepts a wide range of inclination angles, resulting in parameter degeneracy. 
    Another limitation is the assumption of a constant $h/r$ for the height of the scattering surface of the outer disk, which does not consider the radial variation in a real disk. Such variations can significantly change the position of the shadows and thus, the estimation of the shape of the inner disk.
    
    To overcome these limitations, we developed a novel method distinct from previous research in two critical respects. First, it incorporates a finite inner disk thickness. Second, it considers the scattering surface height of the outer disk as a function of the radius. Another significant feature is the fitting of the boundary between the illuminated and shadowed regions on the outer disk using a curve. This enables the generation of a more detailed model that incorporates information on the positions and widths of the shadows.
    The key concept in the proposed method is the outer disk surface horizon, which represents the boundary that separates the visible and invisible sides of the outer disk surface from our perspective, thereby enhancing the accuracy of determining the outer disk surface.
    
    The remainder of this paper is organized as follows.
    A detailed formulation of the proposed method is presented in Section 2. The results of the application of this method to infrared scattered light observations of HD100453 is presented in Section~3. The physical quantities and implications of the geometric structures derived from our findings are discussed (Section 4). Finally, we summarize and present our conclusions (Section 5).

    \begin{figure}[t]
    \centering
    \includegraphics[width=0.9\hsize]{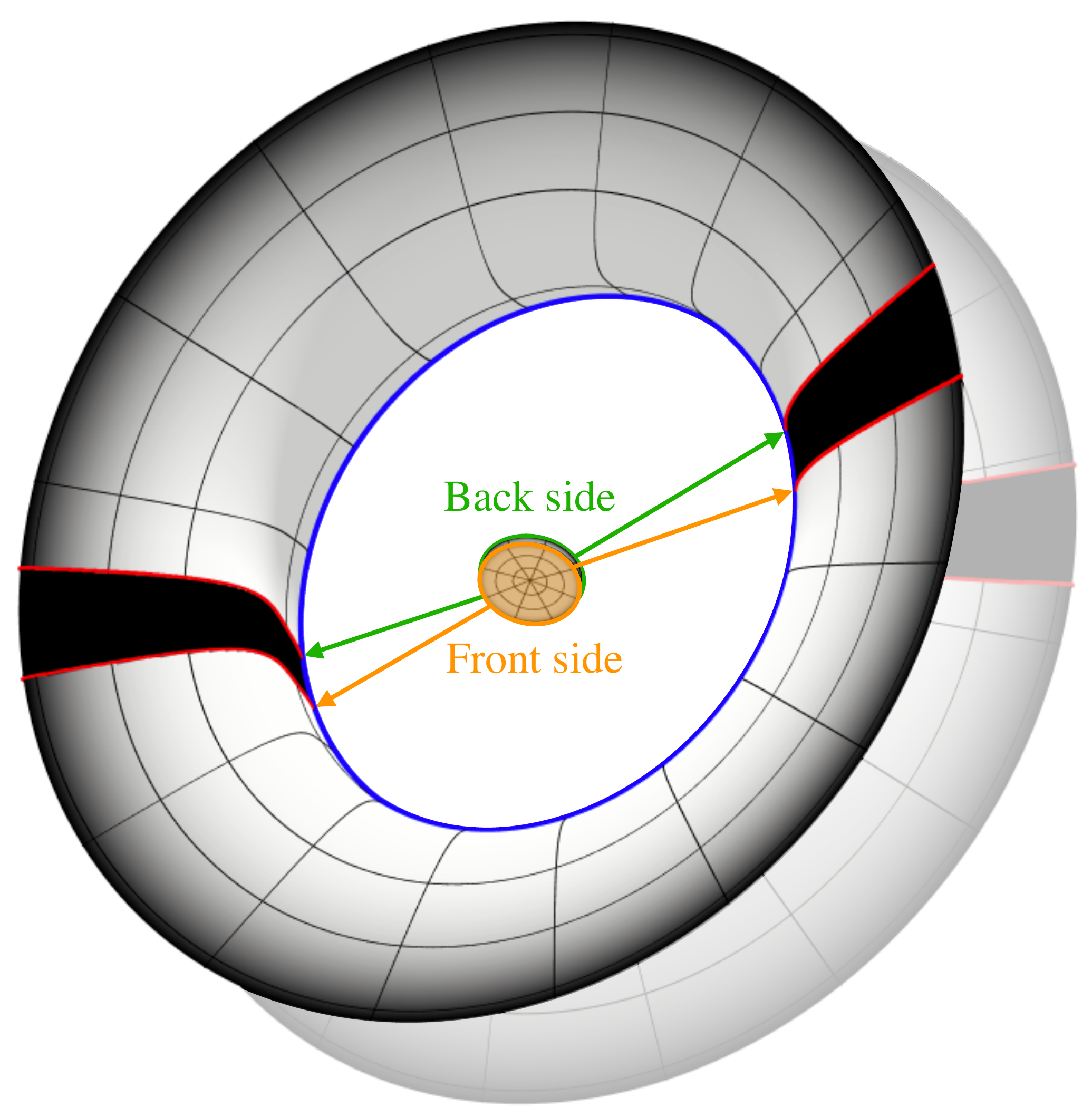}
    \caption{
    Schematic of the scattering surface of the transition disk with the misaligned inner disk.
    The black regions represent shadows cast by the misaligned inner disk, and the red curves indicate the boundary between the illuminated and shadowed areas. The blue curve represents the horizon curve of the outer disk.
    }
    \label{fig:shadow_schematic}
    \end{figure}

\section{Methods} \label{sec:methods}

    This section presents the parameter search methods for exploring the geometric structures of the inner and outer disk surfaces. It provides precise definitions of the horizon and shadow boundaries, followed by a detailed explanation of the optimization techniques employed to refine the disk geometry model based on these curves.
    
    \begin{figure}[t]
    \centering
    \includegraphics[width=\hsize]{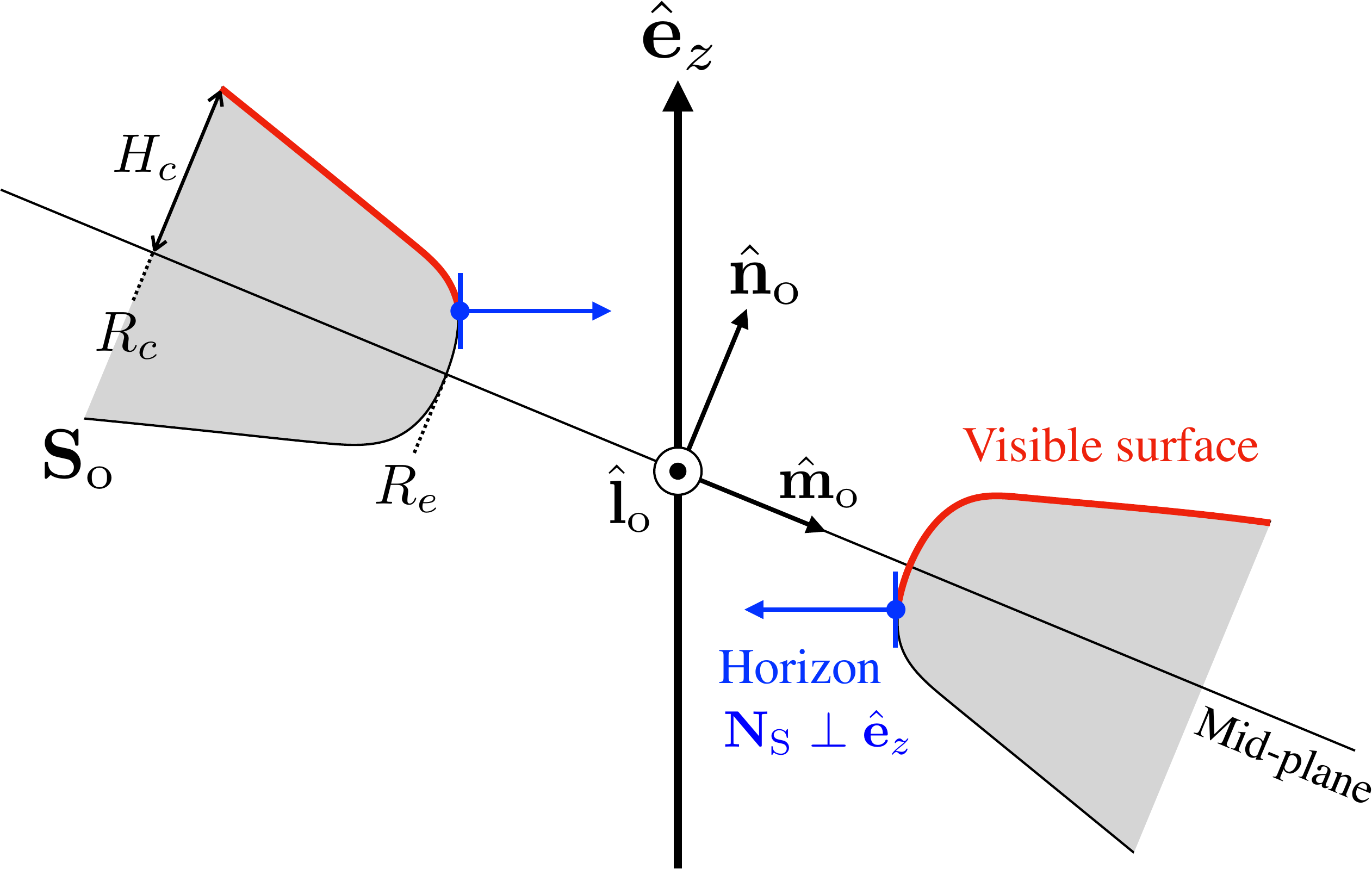}
    \caption{Horizon in the optically thick surface of the outer disk cross-section viewed along the major axis ($\blo$-axis). The surface of the outer disk is modeled such that the radius of the inner edge is $R_e$, and the surface height at a radius of $R_c$ is $H_c$. The position of the horizon, marked by the blue dots, is defined by the condition $\Ns\cdot\bez=0$. The red curve represents the visible surface, and the surface beyond the horizon is not visible to us.}
    \label{fig:horizon}
    \end{figure}

\subsection{Horizon of the Outer Disk Surface}

    The coordinate system is defined such that the positive x-axis points towards the north, positive y-axis towards the east, and positive z-axis towards the observer. In this system, the unit vector $\bez$ represents the line of sight pointing from the celestial sphere towards the observer, whereas the vectors $\bex$ and $\bey$ correspond to the directions of declination (Dec) and right ascension, respectively.
    
    The normal vector of the outer disk midplane is aligned with the direction.
    \begin{equation}
        \bno = 
            \begin{bmatrix}
            -\sin(\io) \sin(\pao) \\
            \sin(\io) \cos(\pao) \\
            \cos(\io)
            \end{bmatrix},
    \end{equation}
    where $\ii$ is the inclination of the outer disk and $\pao$ is the position angle of its major axis (Figure~2).
    In addition to this $\bno$, we define $\blo$ and $\bmo$ as follows as a basis for describing disk geometry:
    \begin{align}
        &\blo = \frac{\bno\times\bez}{|\bno\times\bez|},\\
        &\bmo = \bno\times\blo.
    \end{align}
    Here, $\blo$ denotes the direction parallel to the major axis of the outer disk.

    We assume that the outer disk is optically thick at the wavelengths at which the scattered light is observed. 
    With these basis vectors $\{\blo,\bmo,\bno\}$, the surface of the outer disk $\So$ in Figure \ref{fig:horizon}, centered at $\Co$, can be described as a function of the vertical height $H$ and the azimuthal angle $\phio$:
    \begin{align}
        \So = R(H)(\cos(\phio)\blo + \sin(\phio)\bmo) + H\bno + \Co.
    \end{align}
    The position $\Co = (\Cox,\Coy,\Coz)$ represents the center of the outer disk relative to the image center.
    Based on the offsets $\Cox$ and $\Coy$ on the celestial sphere, the coordinates $\Coz$ can be calculated using the following equation:
    \begin{align}
    \Coz = -\frac{\bno \cdot (\Cox\bex+\Coy\bey)}{\bno\cdot\bez}.
    \end{align}
    $R$ represents the cylindrical radius associated with the height $H$: 
    \begin{align}
        R(H)=\left[R_e^\alpha + (R_c^\alpha - R_e^\alpha)\lr{\frac{|H|}{H_c}}^\alpha\right]^{\frac{1}{\alpha}},
    \end{align}
    where $R_e$ denotes the radius of the inner edge of the outer disk, $R_c$ denotes the characteristic radius, $H_c$ denotes the surface height at $R_c$, and $\alpha$ denotes the power-law index that controls the curvature of the surface as shown in Figure \ref{fig:horizon}.
    Appendix A provides the rationale for selecting this equation.
    To express the curvature of the disk, the derivative of $R(H)$ with respect to $H$ is required. This is expressed as follows:
    \begin{align}
    \frac{dR}{dH} = \sign{H}\frac{R_c^\alpha - R_e^\alpha}{H_c^\alpha} R^{1-\alpha} |H|^{\alpha-1}.
    \end{align}
    
    The surface orientation is characterized by the normal vector $\Ns$, which is obtained by taking the cross product of the partial derivatives of $\So$ with respect to the surface height $H$ and azimuthal angle $\phio$:
    \begin{align}
        \Ns = \deldel{\So}{H} \times \deldel{\So}{\phio},
    \end{align}
    where
    \begin{align}
        &\deldel{\So}{H} =\deldel{R}{H} (\cos(\phio)\blo+\sin(\phio)\bmo)+\bno, \\
        &\deldel{\So}{\phio} = R(-\sin(\phio)\blo + \cos(\phio)\bmo).
    \end{align}
    Using the expressions for the partial derivatives, Equation (8) can be rewritten in the following form:
    \begin{align}
        \Ns = R\left(-\cos(\phio)\blo-\sin(\phio)\bmo+\deldel{R}{H}\bno\right).
    \end{align}
    
    The horizon is the boundary between the visible and invisible sides of the outer disk surface, as seen by the observer in scattered light at the OIR wavelengths, as shown in Figure \ref{fig:horizon}.
    The surface horizon must satisfy the condition that the normal vector of the surface is perpendicular to the line of sight. 
    Therefore, the equation for the horizon is as follows:
    \begin{align}
        \Ns \cdot \bez = 0.
    \end{align}
    Recalling that $\blo$ and $\bez$ are orthogonal, we can simplify this condition as follows:
    \begin{align}
        \phi_{\rm o,Hor} = &\arcsin\left(\deldel{R}{H}\frac{\bno\cdot\bez}{\bmo\cdot\bez}\right) \notag \\
        &{\rm or}\ \pi - \arcsin\left(\deldel{R}{H}\frac{\bno\cdot\bez}{\bmo\cdot\bez}\right)
    \end{align}
    This relationship enables the horizon curve on the disk surface, $\mathbf{C}_{\rm Hor} = \So(H; \phi_{\rm o,Hor}(H))$, to be derived analytically as a function of $H$.
    Using this model significantly tightens the constraints on key parameters such as the central position and radius of the inner edge of the outer disk.

\subsection{Shadow Boundary between the Inner and Outer Disks}

\begin{figure}[t]
    \centering
    \includegraphics[width=\hsize]{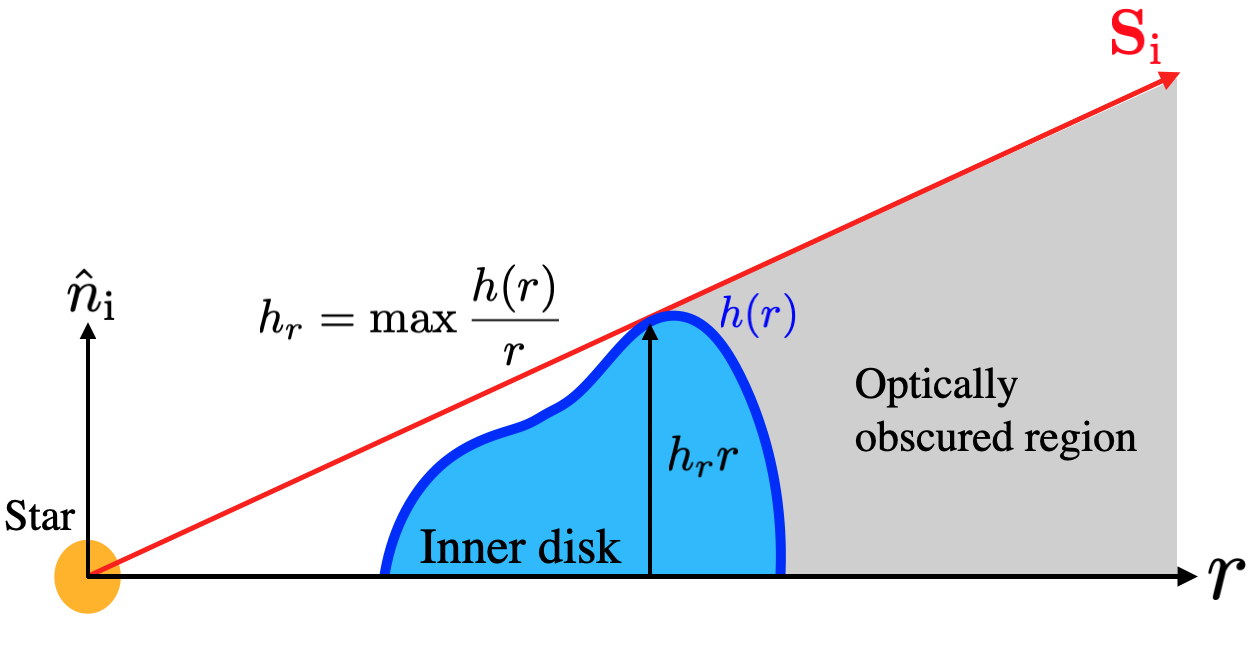}
    \caption{Optically obscured region made by the inner disk.}
    \label{fig:innerdisk}
    \end{figure}

    To capture the three-dimensional shape of the shadow cast by the misaligned inner disk, we modeled the boundary between the illuminated and shadowed regions on the outer disk. 
    As shown in Figure \ref{fig:innerdisk}, the shadow boundary is mathematically defined as the intersection of the surface of the outer disk and the optically obscured region created by the inner disk.
    As a result, the shadow projected onto the outer disk does not capture the surface profile of the inner disk $h(r)$, but provides information about the maximum aspect ratio $h_r = \max h(r)/r$.
    \chtext{In this paper, the term aspect ratio refers to the ratio between the height of the OIR scattering surface and the radial distance from the star. It should be noted that this is not the commonly used definition based on the gas pressure scale height.}
    
    Before formulating the disk model, we redefine the basis vectors for each disk. The normal vector of the inner disk, $\bni$, is defined by the inclination $\ii$ and the position angle $\pai$, as given in Equation~(1).
    These vectors are crucial for computing the misalignment angle between disks, which is a key parameter in our model. The misalignment angle $\Delta \phi_{\rm mis}$ is calculated using the dot product of $\bno$ and $\bni$ as follows:
    \begin{align}
        \Delta \phi_{\rm mis} &= \arccos[\bno \cdot \bni]\\
        &= \arccos\left[ \sin(\ii) \sin(\io) \cos(\pai-\pao)\right. \notag \\
        & \hspace{1.5cm} \left. + \cos(\ii) \cos(\io) \right]. \notag
    \end{align}
    
    Given $\bno$ and $\bni$, the remaining vectors $\blo$, $\bmo$, $\bli$, and $\bmi$ are defined as follows.\footnote{The vectors $\blo$ and $\bmo$ in this subsection are different from those in \S~2.1 with the same notation.}
    \begin{align}
        &\bli = \blo = \frac{\bno\times\bni}{|\bno\times\bni|},\\
        &\bmi = \bni\times\bli,\\
        &\bmo = \bno\times\blo.
    \end{align}
    
    To derive the shadow intersection curve $\mathbf{C}_{\rm Sha}$, the model considers the height profiles of both the inner and outer disks. Using the defined basis vectors $\{\bli,\bmi,\bni\}$, the surface of the optically obscured region caused by the inner disk $\Si$ can be expressed as follows:
    \begin{align}
        \Si = r(\cos(\phii)\bli + \sin(\phii)\bmi) \pm h_r r\bni + \Ci,
    \end{align}
    where $r$, $\phii$, and $\Ci$ denote the cylindrical radius, azimuthal angle in the midplane of the inner disk, and position vector of the inner disk center, respectively. For simplicity, it was assumed that the center of the inner disk coincided with the position of the central star.
    Similar to $\Co$ on the outer disk, given offsets $\Cix$ and $\Ciy$ on the celestial sphere, $\Ciz$ can be computed as follows:
    \begin{align}
        \Ciz = -\frac{\bno \cdot (\Cix\bex+\Ciy\bey)}{\bno\cdot\bez},
    \end{align}
    where we assume that the star is located at the center of the inner disk and lies on the midplane of the outer disk, resulting in the relation $(\Co-\Ci)\cdot\bno=0$. Furthermore, because the system is defined by $\Co\cdot\bno=0$, the relationship $\Ci\cdot\bno=0$ must also hold true.
    The $\pm$ sign in the third term denotes the front and back surfaces of the optically obscured region relative to the line of sight.
    
    The condition for the intersection curve between the surfaces of the outer disk and the optically obscured region created by the inner disk is as follows:
    \begin{align}
        \So-\Si = 0.
    \end{align}
    By calculating this equation, as shown in Appendix B, shadow boundary curves  $\mathbf{C}_{\rm Sha} = \So(H; \phi_{\rm o,Sha}(H))$ can be analytically derived.

    \begin{deluxetable*}{llllll}[t]
    \label{tab:model parameters}
    \tablewidth{1\textwidth} 
    \tablecaption{Model Parameters}
    \tablehead{
    \colhead{Parameter} & \colhead{Description} & \colhead{Search range}& \colhead{Solution} & \colhead{Error}& \colhead{Unit}\\
    \colhead{} & \colhead{} & \colhead{[min, max]} &\colhead{} & \colhead{} & \colhead{}
    }
    \startdata 
    $\Cox$& Offset of the outer disk center in the x-direction (Dec) & [$-2$, 2]& 0.34& 0.016& [au]\\
    $\Coy$& Offset of the outer disk center in the y-direction (RA) & [$-2$, 2]& 0.46& 0.022& [au]\\
    $R_e$& Radius of inner edge of outer disk & [15, 17.5]& 16.8& 0.03& [au]\\
    $H_c$ & Height of the outer disk surface at $R_c=25\ \rm au$ & [2, 12]& 6.6& 0.15& [au]\\
    $\alpha$ & Power index of outer disk surface & [2, 5]& 4.0& 0.17& [ ]\\ 
    $\Cix$& Offset of the inner disk center in the x-direction (Dec) & 0 (fixed)& -& -& [ ]\\
    $\Ciy$& Offset of the inner disk center in the y-direction (RA) & 0 (fixed)& -& -& [ ]\\
    $\ii$ & Inclination of inner disk & [0, 90]& 50.5& 0.8& [$^\circ$]\\
    $\pai$ & Position angle of inner disk & [0, 180]& 88.3& 0.5& [$^\circ$]\\
    $h_r$ & Maximum aspect ratio of inner disk & [0.05, 0.2]& 0.17& 0.001& [ ]\\ \bottomrule
    \enddata
    \end{deluxetable*}

\subsection{Gradients of Infrared Scattered Light Image}

    To estimate the geometric structure of the disk using our model, it was necessary to extract the corresponding quantities for the horizon and boundaries from the image. The highest gradients between adjacent pixel values were utilized because they represent the transitions at the horizon and boundary between the illuminated and shadowed regions.
    
    Boundary extraction from OIR scattered light images is often challenging because of the reduced contrast caused by the propagation characteristics of light within the scattering media. To address this issue, we employ the  contrast-limited adaptive histogram equalization (CLAHE) algorithm \citep{Sasi_2013}. CLAHE is  effective in enhancing the local contrast while mitigating excessive noise amplification. By equalizing the overall contrast of the image, the algorithm is expected to significantly enhance the accuracy of the boundary extraction.
    
    To detect the edges of the scattered light intensity $I_\nu$, the Canny method was employed to compute the gradient vectors \citep{Canny_1986}. The Canny method is a widely used edge detection technique comprising several sequential steps, including Gaussian filtering for noise reduction, gradient calculation, non-maximum suppression, and hysteresis thresholding for binarization. However, in this study, only gradient calculation and non-maximum suppression steps were utilized.
    
    Positions at high gradients were selected, and the directional information of the gradients was further analyzed to separate the horizon and shadow components. Specifically, the computed gradient vectors were projected onto the basis vectors in radial and azimuthal directions, $\hat{\bf{e}}_r$ and $\hat{\bf{e}}_\phi$, by using inner product calculations. Consequently, the $\hat{\bf{e}}_r$ component of the gradient vector corresponds to the horizon, whereas the $\hat{\bf{e}}_\phi$ component corresponds to the shadow edge.
    
    However, it is important to note that  the computation of gradients may also capture small-scale structures other than shadows, such as spirals. To address this potential ambiguity, a mask is applied to constrain the shadow on the outer disk.

\subsection{Parameter Exploration}

    We generated horizon and shadow boundary curves to determine the parameters that minimized the distance between these curves and the fast gradient points. The parameters obtained from this modeling process are listed in Table 1. For the outer disk, the parameters are $\Cox$, $\Coy$, $R_e$, $H_c$, and $\alpha$. The inner disk includes $\Cox$, $\Coy$, $\ii$, $\pai$, and $h_r$. 
    The inclination and position angle of the outer disk were determined based on the dust continuum emission and CO channel maps observed by the ALMA.
    
    To efficiently explore the parameter space and estimate uncertainties, we employed the Markov chain Monte Carlo (MCMC) method \citep{Foreman-Mackey_2013}. To enhance both the efficiency and diversity of MCMC sampling, we utilized parallel tempering, in which multiple chains with different "temperatures" are run simultaneously, and samples are periodically exchanged between these chains. This method allows chains to escape from local minima, facilitating a broader exploration of the parameter space and enabling the identification of global minima, which are often challenging to capture using conventional MCMC methods.
    
    For the prior distributions of each parameter, we adopt uninformative priors. This approach considers information derived from observational data by minimizing the influence of prior assumptions. The specific prior settings for each parameter are listed in Table~1.
    
    \begin{figure*}[hbtp]
    \centering
    \includegraphics[width=\hsize]{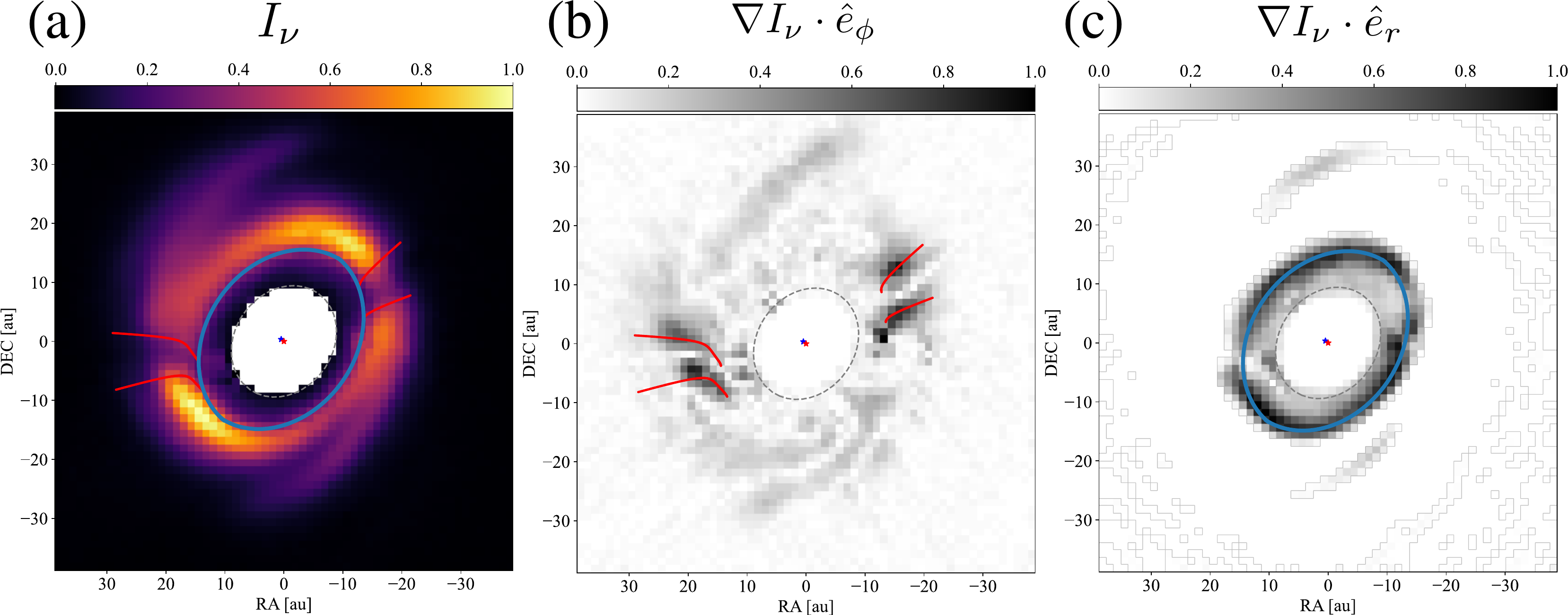}
    \caption{The (a), (b), and (c) panels show the infrared scattered light image, azimuthal gradient image, and positive radial gradient image, respectively. All images are displayed in arbitrary units. The blue and red curves represent the optimized horizon curve and shadow intersection curve, respectively. The blue and red star markers at the center indicate \chtext{the centers of the outer and inner disk, respectively.}}
    \label{fig:shadow_results}
    \end{figure*}

\section{Application to Real Data: HD100453} \label{sec:results}

    By adopting the model presented in Section~2, the disks around HD 100453, located at a distance of 103.78~pc \citep{Gaia_2021}, are modeled from the OIR scattered light image. The image was observed using the high-contrast imaging instrument spectro-polarimetric high-contrast exoplanet research (SPHERE) mounted on the Very Large Telescope (VLT) in the $J$-band under program ID 096.C-0248(B) and was published by \cite{Benisty_2017}.

    Observations were made with a polarized intensity. If there is a significant variation in the degree of polarization across the image, care should be taken when determining the shadow positions. This is particularly important for disks with large tilt angles. However, in this case, the presence of shadows was too extreme to be explained solely by a decrease in the degree of polarization. Therefore, we assume that the effect of the polarization degree is negligible. 
    
    The position of the star (center of the inner disk) was determined with high accuracy using satellite spots obtained using the waffle mode of the VLT/SPHERE. During the data reduction process, the star was aligned to the center of the image; therefore, we fixed both $\Cix$ and $\Ciy$ at zero.
    In addition, PA and $i$ of the outer disk were fixed at $324^\circ$ and $34^\circ$, respectively, based on \cite{Bohn_2022}.
    
    Figure~\ref{fig:corner} in Appendix C shows the posterior distributions obtained from MCMC fitting. The parameters with maximum posterior probabilities derived from these distributions are listed in Table~1. 
    Figure \ref{fig:shadow_results} shows the optimized curves generated using the proposed method. In panel~(b), the shadow curve is overlaid onto the azimuthal gradient distribution, and in panel~(c),  the horizon is overlaid onto the radial gradient distribution. These curves were used to trace the distributions of intensity gradients. 
    Compared to the study by \cite{Min_2017}, our study presents new findings in that we obtained the surface heights of both the inner and outer disks from scattered light images.
    
    In the following, we describe in detail the geometric structure of HD 100453 derived from the optimization.
    
\subsection{Geometry of the Inner Disk}

    The PA and inclination ($i$) of the inner disk were estimated to be $88^\circ$ and $50^\circ$, respectively. These results are generally consistent with the measurements of the inclination of the inner disk obtained using VLTI/GRAVITY \citep{Bohn_2022}, demonstrating the robustness of this orientation. Furthermore, the misalignment angle of the inner disk relative to the outer disk, as described by Equation~(14), was estimated to be $\sim70^\circ$. According to \cite{Nealon_2020}, to generate such a large tilt, an additional planetary companion of $\sim 5 M_{\rm J}$ is required inside the outer disk, 
    \chtext{along with a known stellar companion of $\sim 0.2 M_{\odot}$ \citep{Chen_2006}.
    \cite{Rosotti_2020} suggest that the localized features identified in the line-of-sight velocity map could be caused by the planet embedded in the disk. However, they note that the resolution of their data is insufficient to confirm this hypothesis, and that follow-up observations with higher resolution are needed.
    }

    \chtext{
    The maximum aspect ratio $h_r$ of the inner disk was estimated to be $0.17$.
    Assuming that the star with a luminosity of $6~L_\odot$ \citep{Bohn_2022} hosts an optically thin inner disk, the gas pressure scale height at a radius of 1~au is estimated to be approximately 0.03~au.
    The maximum aspect ratio $h_r$ of the inner disk can be more than five times higher than the gas pressure scale height.
    This suggests that the number density of small dust particles above the pressure scale height is so high that the dust scattering surface is located higher than the pressure scale height. It is possible that the dust tightly coupled to the gas is vertically dispersed by the gas turbulence or disk wind. 
    }

    \begin{figure*}[t]
    \centering
    \includegraphics[width=\hsize]{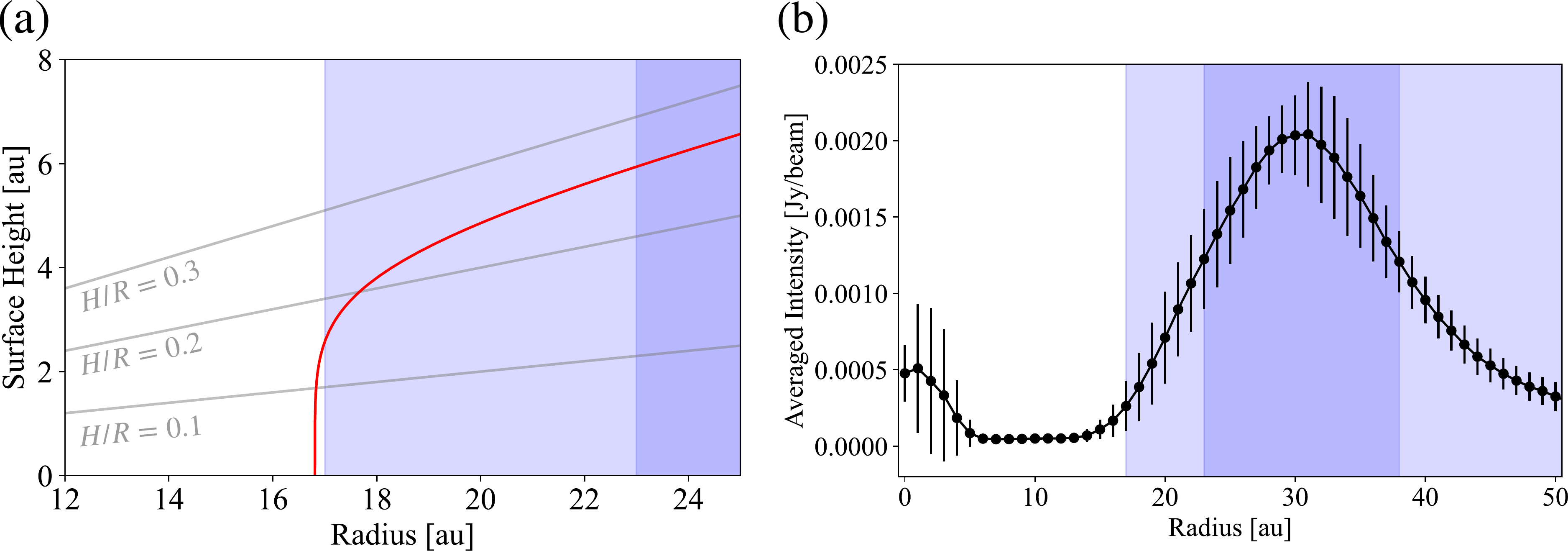}
    \caption{(a) Estimated disk surface height profile represented by Equation~(6) (red line). 
    (b) The azimuthally averaged intensity profile of the dust continuum emission observed with ALMA Band~7 \citep{Rosotti_2020}. The error bars represent the standard deviation of the intensity at each radius. The dark blue shaded region indicates areas with an intensity greater than half of the peak intensity, while the light blue shaded region corresponds to intensities greater than one-tenth of the peak intensity. The shaded regions in panel (a) follow the same definitions.}
    \label{fig:surface_profile}
    \end{figure*}  

\subsection{Surface Height Profile of the Outer Disk}

    The location of the scattering surface of the outer disk is presented in Figure. \ref{fig:surface_profile}, and it is evident that the aspect ratio of the outer disk ranges from 0.15 to 0.25. Parameter $\alpha$ was estimated to be $\sim 4$, and the height profile converged rapidly to form $H \propto R$. 
    In addition, as shown in Figure~\ref{fig:corner}, $H_c$ and $\alpha$ exhibit negative correlations. When $H_c$ is low, $\alpha$ is increased to maintain the surface at a relatively high position. Conversely, when $H_c$ is high, $\alpha$ decreases but maintains the surface at a similar height. The interplay between $H_c$ and $\alpha$ strongly supports the presence of an optically thick dusty surface at high positions near the inner edge of the outer disk.

    As illustrated in Figure \ref{fig:surface_profile} (b), the azimuthally averaged intensity of the dust continuum emission observed in ALMA Band~7 \citep{Rosotti_2020} exhibits an increase of approximately 17~au and a peak near 30~au. The 17~au corresponds to the radius of the inner edge of the outer disk identified in OIR scattered light, whereas the high density of millimeter-sized dust is observed further out, around 30~au.These comparisons suggest that larger dust grains are more likely to be trapped at pressure bumps, while smaller dust grains drift inward to regions closer than this location.

\section{Discussion} \label{sec:discussion} 

\subsection{Strengths of our Model}

    Our new approach successfully constrained the geometric structures of both the outer and inner disks in more detail by modeling the horizon and shadow boundaries.
    
    The determination of the horizon is particularly important for constraining the central position $(\Cox,\Coy)$ and inner edge radius $R_e$ of the outer disk. The horizon exhibited an elliptical shape, and the orientation and length of its major axis were dependent on the central position but not strongly correlated with the surface height of the outer disk.  
    By contrast, the vertical location of the scattering surface of the outer disk ($H_c$, $\alpha$) influences the length of the minor axis of the horizon. However, this dependency is very small, making it difficult to determine these parameters precisely. Fortunately, these parameters have a significant impact on the shapes of the shadow boundaries, enabling better constraints through the modeling of shadow boundaries.
    
    Earlier studies, particularly shadow analysis by \cite{Min_2017}, used a shadow model that did not consider the thickness of the inner disk. The model calculates the width of the shadow using the PA and $i$ values of the inner disk and the fixed aspect ratio of the outer disk. However, this leads to problems with degeneracy in the combination of the PA and $i$ values.
    To improve this, we incorporated the aspect ratio of the inner disk into the model and integrated the shadow width information. This improvement resolved the degeneracy of the PA and $i$ of the inner disk, thereby enabling a more accurate estimation of the disk structure.

\chtext{
\subsection{Flared Disk}

    In the present model, the disk height converged to $H\propto R$ for large radii. However, some disks have flared shapes. This implies that the model can be improved for use in these situations. The formula for the location of the scattering surface is as follows:
    \begin{align}
    &R(H)=\left[R_e^\alpha + (R_c^\alpha - R_e^\alpha)\lr{\frac{|H|}{H_c}}^{\frac{\alpha}{\beta}}\right]^{\frac{1}{\alpha}},\\
    &\frac{dR}{dH} = \sign{H}\frac{R_c^\alpha - R_e^\alpha}{\beta H_c^{\frac{\alpha}{\beta}}} R^{1-\alpha} |H|^{\frac{\alpha}{\beta}-1}.
    \end{align}
    In these equations, the disk surface asymptotically approaches $H \propto R^\beta$ in the outer region. Parameter $\alpha$ adjusts the asymptotic rate in the radial direction, whereas $\beta$ controls the flaring index.
    
    Because the parameters $\alpha$ and $\beta$ are correlated, it can be difficult to determine them accurately.
    To mitigate this problem, including an apparent outer edge of the outer disk in the model can improve the accuracy of the parameter estimation. Here, the term apparent refers to the observed edge of the outer disk in scattered light, which may not coincide with the true physical outer edge. This is because the true outer edge could be further out, but be hidden behind the shadow cast by the disk surface located at smaller radii within the disk.

    In this framework, the apparent outer edge radius $R_{\rm{out}}$ is treated as a free parameter and is expressed as follows:
    \begin{align}
        \mathbf{C}_{\rm out} &= \So(\phio; R_{\rm out}) \notag \\
        &= R_{\rm out}(\cos(\phio)\blo + \sin(\phio)\bmo) + H_{\rm out}\bno + \Co,
    \end{align}
    where $H_{\rm out}$ is the surface height of the outer disk with radius $R_{\rm out}$.
    Care must be taken when applying this model to a disk with significant non-axisymmetric structure.

    We applied this model to the OIR scattered light image of HD 100453 and found results consistent with those obtained without considering flaring.
    Figure~\ref{fig:shadow_results_withflaring} in Appendix D shows the optimized curves obtained from the fitting procedure with the flared disk model. 
    The locations of the apparent outer edge were determined from the distribution of the absolute values of the negative components of the radial gradient, using only regions along the disk's major axis of the disk where the influence of spiral arms appears to be minimal. 
    
    Figure~\ref{fig:corner_withflaring} in Appendix D shows the posterior distributions from the MCMC fit.
    The parameters with maximum posterior probabilities derived from these distributions are listed in Table~2. 
    For the inner disk parameters, the results are consistent with those obtained by the model presented in section~3. For the outer disk, on the other hand, the best-fit flaring index $\beta$ is found to be very close to 1, suggesting that the disk is not significantly flared within the region traced by the scattered light.
    It should be noted, however, that the possibility of flaring beyond $R_{\rm{out}}$ cannot be ruled out.
}

\subsection{Limitations of our Model and its Improvement Strategies}

    The proposed model assumes that the vertical structure of the inner disk is symmetric with respect to the midplane. However, global magnetohydrodynamic (MHD) simulations by \cite{Bai_2017} showed that asymmetric disk winds can develop on both sides of the disk. 
    Therefore, the vertical structures of the upper and lower layers may differ, and the inner disk may cast more complex shadows on the outer disk. In light of this, the current model may not fully reproduce the shapes and positions of shadows, implying a need for future refinement. To address such asymmetries, it may be necessary to use three-dimensional radiative transfer simulations, enabling comparisons with observations not only in terms of geometry but also in terms of intensity.

\subsection{Outlook Based on the New Information Provided by our Method}

    The modeling approach developed in this study, along with the obtained results, provides a foundation for further analysis and discussion. This section discusses possible future research directions and physical mechanisms that may be explored. 
    
    \subsubsection{Further Constraints on Inner Disk Geometry}
    
    The information estimated from the proposed model will facilitate a more precise characterization of dust properties through SED analysis. By employing the orientation ($\ii,\pai$) and aspect ratio $h_r$ of the inner disk as fixed parameters in the radiative transfer calculation, comparisons of the observed SED enable a detailed analysis of the vertical structure of dust density, size, and temperature within the inner disk.

    \chtext{
    According to \cite{Flock_2016}, the aspect ratio reaches its maximum value at the outer edge of the silicate sublimation front, which corresponds to the inner edge of the dead zone \citep{Ueda_2017}.
    In the simulations by \citet{Flock_2016}, the aspect ratio of the gas pressure scale height at the inner edge of the dead zone in disks around Herbig Ae/Be stars is in the range of $0.09-0.14$. Our derived $h_r$ appears somewhat higher in comparison.
    }

    On the other hand, if dust coupled with gas is blown away by the MHD wind and becomes optically thick \citep{Miyake_2016}, it may affect the shadow width. Distinguishing between the actual width of the inner disk and the influence of wind is challenging, requiring multi-epoch observations, such as simultaneous measurements of the velocity field of the inner disk and variations in the shadow shape.

\subsubsection{Scattering Angles and Phase Function on the Outer Disk Surface}

\begin{figure}[t]
    \centering
    \includegraphics[width=0.9\hsize]{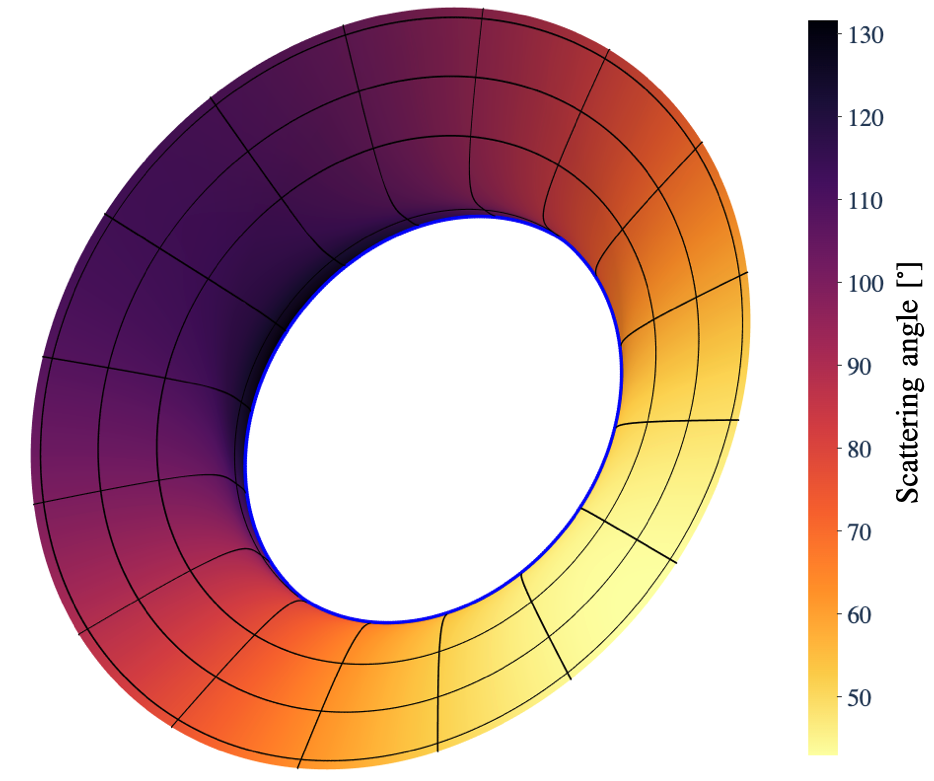}
    \caption{Scattering angle of the dust on the outer disk surface.}
    \label{fig:scatteringangle}
    \end{figure}

    The capacity of the model to precisely estimate the stellar position and surface height distribution of the outer disk facilitates the determination of the scattering angle at each position, as illustrated in Figure \ref{fig:scatteringangle}. The scattering angle is calculated as follows: 
    \begin{align}
        \theta_{\rm scat} = \arccos\left(\frac{\So-\Ci}{|\So-\Ci|}\cdot\bez \right).
    \end{align}
    By comparing this scattering angle with the total intensity, we derived a dust phase function \citep{Ginski_2023}. The phase function represents the scattering characteristics of light by dust particles and allows us to estimate the size, shape, and composition of the dust from the observed scattered light.
    
    In Section~3, we used the polarized intensity for analysis. Further observational studies are necessary to advance this direction, particularly in the imaging of the total intensity, although this is technically demanding.

\subsubsection{Measuring the Cooling Time of Material in the Outer Disk}

    Heating from the central star is reduced in the shadow, leading to a temperature drop caused by the cooling of dust particles \citep{Casassus_2019}. By comparing the three-dimensional spatial distribution of shadows estimated from the present model with the temperature distribution of the disk derived from submillimeter observations using ALMA, it is possible to measure the cooling time of the dust. Notably, if the cooling time is significantly shorter than the Keplerian timescale, local temperature gradients may trigger fluid instabilities such as vertical shear instability (VSI) \citep{Flock_2020, Fukuhara_2023}. VSI is thought to be one of the mechanisms driving turbulence, which is expected to play a critical role in planetesimal formation by promoting dust growth.
    
    According to multi-epoch observations by \cite{Pinilla_2018}, it has been confirmed that the shadows within the disk associated with J1604 exhibit variations on short timescales.
    A comparison of OIR images with submillimeter images in different epochs involves the risk of misaligning the actual spatial relationship between shadows and temperature distribution. Such a misalignment can result in incorrect estimations of cooling times.
    To address this issue in the context of cooling time measurements, it is necessary to select objects that exhibit sufficiently long shadow variability timescales or conduct simultaneous observations at OIR and submillimeter wavelengths.

\section{Conclusions} \label{sec:conlusion}
    
    This study proposes a novel method for inferring the three-dimensional geometries of a transition disk system comprising inner and outer disks. This method analyzes the shadows cast by misaligned inner disks onto outer disks and the horizon on the outer disk surface in OIR scattered light images. 
    
    The proposed method offers several improvements over previous approaches, enabling the extraction of critical information regarding disk systems.
    This allows for the determination of the orientation ($\ii$, $\pai$) and maximum aspect ratio $h_r$ of the inner disks as well as the surface height profile of the outer disk ($R_e$, $H_c$, $\alpha$). Additionally, the method resolves parameter degeneracies that were challenging in earlier methods, particularly the interdependence between the orientation of the inner disk and thickness of the outer disk.

    As a proof-of-concept, the application of this method to the HD 100453 system is also presented, providing significant insights into the geometrical and physical characteristics of its inner and outer disks.
    The inner disk was determined to have an inclination of approximately $50^\circ$ and a position angle of $88^\circ$, resulting in a misalignment angle of $\sim70^\circ$ with respect to the outer disk. This substantial tilt highlights the potential influence of an additional companion in the disk system, as suggested in prior studies. \chtext{The maximum aspect ratio of the inner disk, $0.17$, suggests the presence of the surface height of the optically thick dusty component due to vertical lofting by MHD winds or turbulence.}
    The height profile of the outer disk, by contrast, was characterized by rapid convergence, indicative of dust accumulation influenced by gas pressure maxima.
    
\clearpage
\section*{Acknowledgments}
This work used the SPHERE Data Centre, jointly operated by OSUG/IPAG (Grenoble), PYTHEAS/LAM/CESAM (Marseille), OCA/Lagrange (Nice), Observatoire de Paris/LESIA (Paris), and Observatoire de Lyon.
We thank Myriam Benisty for providing us with access to the $J$-band data from \cite{Benisty_2017}.
To obtain the scientific results presented in this paper, we used the Python programming language \citep{python}, NumPy \citep{Oliphant_2006}, Matplotlib \citep{matplotlib_2007}, emcee \citep{Foreman-Mackey_2013}, and astropy \citep{astropy_2013, astropy_2018} packages.

%



\appendix
\section{Surface Height Profile of The Outer Disk}

In this study, the surface height profile of the outer disk was not derived from first-principles or physical models. Instead, it was represented by an empirical function adopted for this analysis, given by the following equation:
\begin{align}
    H(R) = H_c\lr{\frac{R^\alpha - R_e^\alpha}{R_c^\alpha - R_e^\alpha}}^{\frac{1}{\alpha}}.
\end{align}
This function is selected such that it asymptotically approaches $H\propto R$ at sufficiently large distances from the disk center ($R\gg R_e$). 
The parameter $\alpha$ plays a crucial role in determining the rate at which the function approaches asymptotic behavior. When $\alpha=2$, the function assumes the form of a hyperbolic curve. Specifically, larger values of $\alpha$ causes the function to converge to its asymptotic form more rapidly, whereas smaller values require a longer scale for the function to attain its asymptotic form. 
To calculate the horizon, Equation~(A1) was rearranged into the inverse function $H$, $R(H)$, as shown in Equation~(6).

This definition is also applied to the shadow boundary model described in Section~2.2, and is considered a sufficient approximation in the relatively inner regions where the shadow is visible.
For example, if the aspect ratio of the outer disk surface has a convex maximum at a certain radius, the regions beyond that point will not receive light; consequently, no scattered light will be observable.

\section{Solutions of the Shadow Boundary Curves}

The shadow boundary curves are obtained by solving the following equation with respect to $\phio$, $\phii$, $r$:

\begin{align}
    \So-\Si= R(H)(\cos(\phio)\blo + \sin(\phio)\bmo) + H\bno -r(\cos(\phii)\bli + \sin(\phii)\bmi) \mp h_r r\bni + \dc =0.
\end{align}
This equation can be expressed using the orthonormal bases $\bli$, $\bmi$, and $\bni$, which represent the inner disk, as follows:
\begin{align}
    \So-\Si
    & = (R(H)\cos(\phio) -  r\cos(\phii) + \dc\cdot\bli)\ \bli \notag \\ 
    &\quad + (R(H)\sin(\phio)\bmo\cdot\bmi + H\bno\cdot\bmi -r\sin(\phii) + \dc\cdot\bmi)\ \bmi \notag \\ 
    &\quad+ (R(H)\sin(\phio)\bni\cdot\bmo + H\bno\cdot\bni \mp h_r r + \dc\cdot\bni)\ \bni \notag\\
    &=0.
\end{align}

By combining the three equations through inner product calculations, we can eliminate variables $r$ and $\phii$, ultimately deriving a quartic equation that depends solely on $x = \sin(\phio)$. This equation takes the following general form:
\begin{align}
a_4 x^4 + a_3 x^3 + a_2 x^2 + a_1 x + a_0 = 0,
\end{align}
where $a_4, a_3, a_2, a_1,$ and $a_0$ are coefficients determined by the parameters in the original equations. These coefficients encapsulate the relationships between the geometric and physical parameters of the system, and their specific forms are given by
\begin{align}
    a_4 &= b_2^2,\\
    a_3 &= 2 b_2 b_1,\\
    a_2 &= b_1^2 + 2 b_2 b_0 + 4 R^2 (\dc\cdot\bli)^2,\\
    a_1 &= 2 b_1 b_0,\\
    a_0 &= b_0^2 - 4 R^2 (\dc\cdot\bli)^2.
\end{align}
The intermediate variables $b_2,\ b_1$ and $b_0$ are expressed in terms of the original parameters as follows:
\begin{align}
    b_2 &= R^2 \lr{\lr{\bni\cdot\bno}^2 - \lr{\frac{\bmi\cdot\bno}{h_r}}^2 -1},\\
    b_1 &= 2R\mathbf{H}_\delta\cdot\lr{\lr{\bni\cdot\bno}\bmi + \frac{\bmi\cdot\bno}{h_r^2}\bni},\\
    b_0 &= R^2 + (\dc\cdot\bli)^2 + (\mathbf{H}_\delta\cdot\bmi)^2 - \lr{\frac{\mathbf{H}_\delta\cdot\bmi}{h_r}}^2,
\end{align}
where $\mathbf{H}_\delta=H\bno+\dc$.

The quartic equation yields four solutions, $\mathbf{C}_{\rm Sha} = \So(H; \phi_{\rm o,Sha}(H))$, which can be classified using Equation~(B3). Specifically, the sign of $\cos(\phio)$ distinguishes the two shadows, whereas the sign preceding $h_r$ determines whether the boundary is formed by the front or back surface of the inner disk (see Figure~\ref{fig:shadow_schematic}).

\section{Corner Plot of the MCMC Results}

\begin{figure*}[h]
\centering
\includegraphics[width=\hsize]{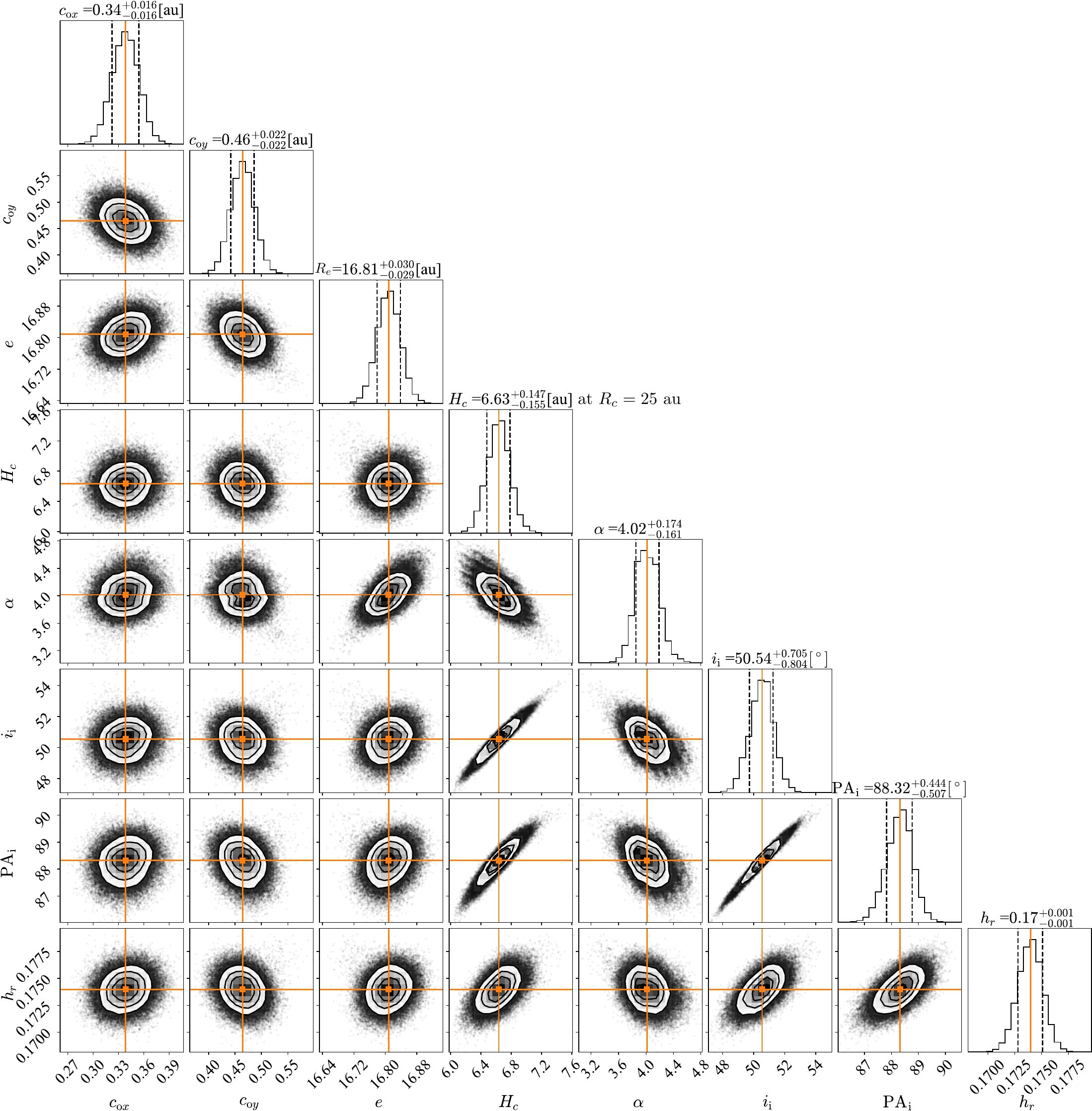}
\caption{MCMC corner plot represents the results of our model fitting. The values and uncertainties shown in the panel titles correspond to the median and the 16th and 84th percentiles, respectively.}
\label{fig:corner}
\end{figure*}

Figure \ref{fig:corner} shows the corner plots of the resultant marginal posterior distributions.

\begin{figure*}[t]
\centering
\includegraphics[width=0.95\hsize]{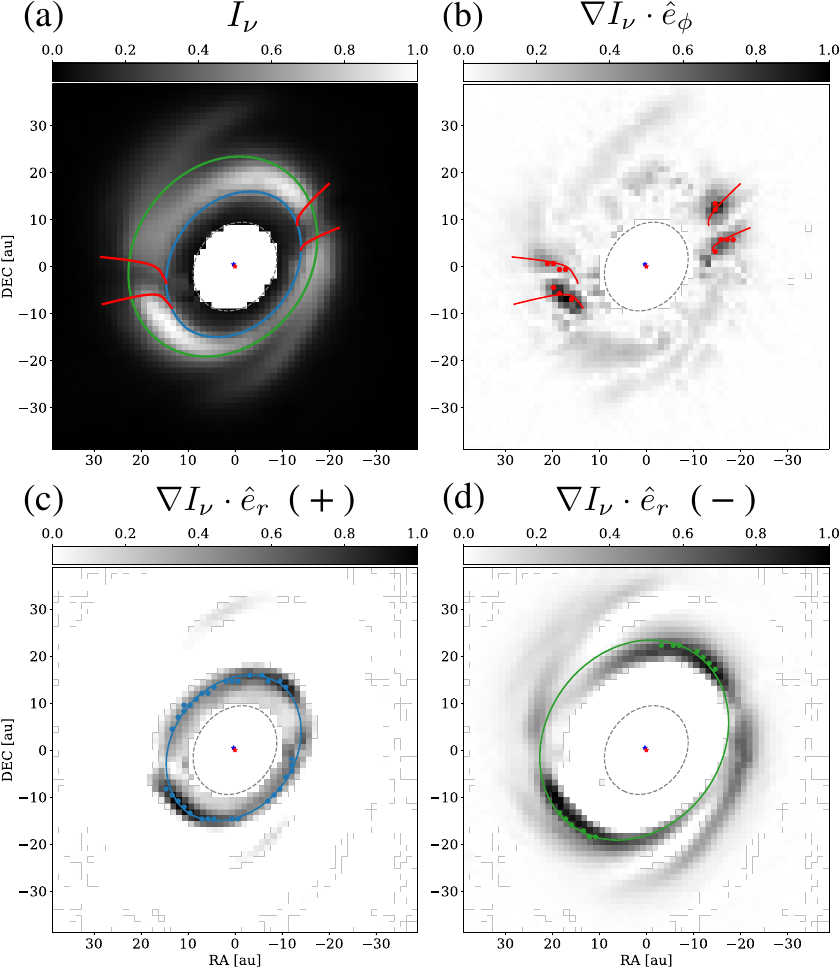}
\caption{The (a), (b), (c), (d) panels show the infrared scattered light image, azimutal gradient image, positive radial gradient image, and negative radial gradient image, respectively. All images are displayed in arbitrary units. The red, blue, and green curves represent the optimized shadow intersection curve, horizon curve, and apparent outer edge curve, respectively. The points in the figures are sampled from each gradient distribution. The blue and red star markers at the center indicate the (geometric) centers of the outer and inner disk, respectively.}
\label{fig:shadow_results_withflaring}
\end{figure*}

\begin{figure*}[t]
\centering
\includegraphics[width=\hsize]{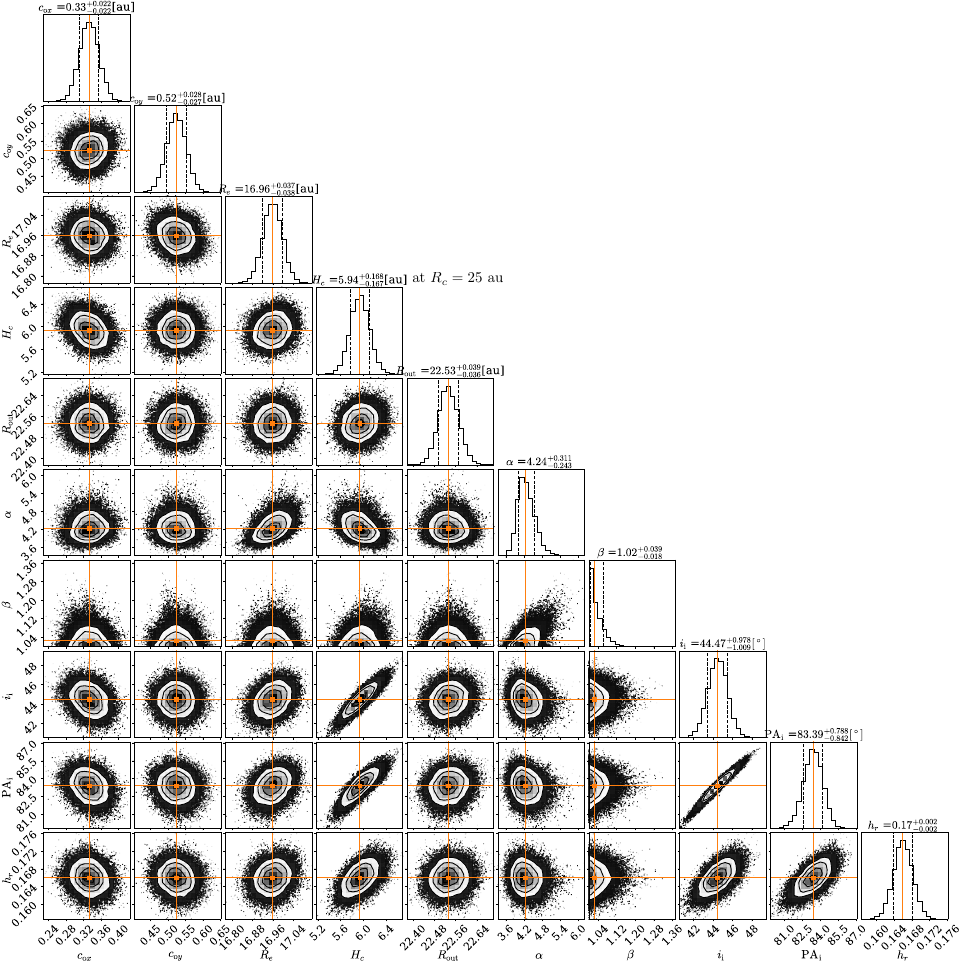}
\caption{MCMC corner plot represents the results of our flared disk model fitting. The values and uncertainties shown in the panel titles correspond to the median and the 16th and 84th percentiles, respectively.}
\label{fig:corner_withflaring}
\end{figure*}

\chtext{
\section{MCMC Results Using the Flared Disk Model}
This section presents the results of the MCMC fitting procedure using the flared disk model. Table~\ref{tab:flaring_params} summarizes the best-fit parameters. Figure~\ref{fig:shadow_results_withflaring} shows the model curves overlaid on the observed images, and Figure~\ref{fig:corner_withflaring} displays the corner plot of the posterior distributions for the fitted parameters.
}

    \begin{deluxetable*}{llllll}[t]
    \label{tab:flaring_params}
    \tablewidth{1\textwidth} 
    \tablecaption{Flared Disk Model Parameters}
    \tablehead{
    \colhead{Parameter} & \colhead{Description} & \colhead{Search range}& \colhead{Solution} & \colhead{Error}& \colhead{Unit}\\
    \colhead{} & \colhead{} & \colhead{[min, max]} &\colhead{} & \colhead{} & \colhead{}
    }
    \startdata 
    $\Cox$& Offset of the outer disk center in the x-direction (Dec) & [$-2$, 2]& 0.33& 0.022& [au]\\
    $\Coy$& Offset of the outer disk center in the y-direction (RA) & [$-2$, 2]& 0.52& 0.028& [au]\\
    $R_e$& Radius of inner edge of outer disk & [15, 17.5]& 16.96& 0.037& [au]\\
    $H_c$ & Height of the outer disk surface at $R_c=25\ \rm au$ & [2, 12]& 6.6& 0.168& [au]\\
    $R_{\rm{out}}$ & Apparent outer edge of the outer disk & [20, 30]& 22.5& 0.39& [au]\\
    $\alpha$ & Power index of outer disk surface & [2, 5]& 4.0& 0.17& [ ]\\ 
    $\beta$ & Flaring index of outer disk surface & [1, 2]& 1.02& 0.039& [ ]\\ 
    $\Cix$& Offset of the inner disk center in the x-direction (Dec) & 0 (fixed)& -& -& [ ]\\
    $\Ciy$& Offset of the inner disk center in the y-direction (RA) & 0 (fixed)& -& -& [ ]\\
    $\ii$ & Inclination of inner disk & [0, 90]& 44.47& 1.0& [$^\circ$]\\
    $\pai$ & Position angle of inner disk & [0, 180]& 83.4& 0.8& [$^\circ$]\\
    $h_r$ & Maximum aspect ratio of inner disk & [0.05, 0.2]& 0.17& 0.002& [ ]\\ \bottomrule
    \enddata
    \end{deluxetable*}

\clearpage
\bibliography{bibliography}{}
\bibliographystyle{aasjournal}



\end{document}